\documentclass[twoside]{dis09}
\usepackage[latin1]{inputenc}
\usepackage[dvips]{graphicx,epsfig,color}
\usepackage{wrapfig,rotating}
\usepackage{amssymb,amsmath,array}

\newcommand\f[2]{\frac{#1}{#2}} 
\newcommand\as{\alpha_{\mathrm{S}}}

\def\tL{{\tilde L}} 
\begin{document}
\title{Transverse-momentum resummation at hadron colliders\footnote{Invited talk given at the XVII International Workshop on Deep-Inelastic Scattering and Related Subjects DIS 2009, 26-30 April 2009, Madrid}}

%***********************************************************************
% AUTHORS INFORMATION AREA
%***********************************************************************
\author{Massimiliano Grazzini
%
% Optional short acknowledgment: remove next line if non-needed
%\thanks{}
%
% DO NOT MODIFY THE FOLLOWING '\vspace' ARGUMENT
\vspace{.3cm}\\
%
% Addresses and institutions (remove "1- " in case of a single institution)
INFN, Sezione di Firenze and
Dipartimento di Fisica, Universit\`a di Firenze,\\
I-50019 Sesto Fiorentino, Florence, Italy\\
%
% Remove the next three lines in case of a single institution
}
%***********************************************************************
% END OF AUTHORS INFORMATION AREA
%***********************************************************************

\maketitle

\begin{abstract}
We consider the transverse-momentum distribution of colourless high-mass systems (lepton pairs, vector bosons, Higgs particles...) produced in hadronic collisions. We briefly review a formalism for the all-order resummation of the logarithmically-enhanced contributions at small transverse momenta, and we present an illustrative selection of numerical results obtained by using our method.

\end{abstract}

The inclusive production of colourless high-mass systems in hadron collisions is important for physics studies within and beyond the standard model.
Well know examples of these processes are the production of lepton pairs, vector boson pairs, Higgs particles and so forth.
These processes are either benchmark processes, whose expected rates should be precisely known,
or irreducible backgrounds for new physics searches.
As such, it is important to have a good theoretical control for the corresponding cross sections and associated distributions.

Among the various distributions, an important role is played by the transverse-momentum spectrum of the high-mass system.

Let us consider the inclusive hard-scattering reaction
\begin{equation}
h_1+h_2\to F(M,q_T)+X,
\end{equation}
where the collision of the two hadrons $h_1$ and $h_2$
produces the triggered final state $F$. The final 
state $F$ consists of one or more colourless particles (leptons, photons, vector
bosons, Higgs bosons, $\dots$) with total invariant mass $M$
and transverse momentum $q_T$.
Note that, since $F$ is colourless, the LO partonic subprocess
is either $q{\bar q}$ annihilation, as in the case of the Drell--Yan process, 
or $gg$ fusion,
as in the case of Higgs boson production.

When considering the transverse-momentum spectrum, it is important to distinguish two regions of transverse momenta.

In the large-$q_T$ region ($q_T \sim M$),
the perturbative series is controlled by a small expansion
parameter, $\as(M)$, and calculations based on the truncation of the series
at a fixed order in $\as$ are theoretically justified.
In the small $q_T$ region ($q_T\ll M$), large logarithmic contributions appear
that spoil the convergence of the ordinary fixed-order expansion and need be
resummed to all orders.

The method to systematically perform all-order resummation of
classes of logarithmically-enhanced terms at small $q_T$ is known
\cite{Dokshitzer:hw}--\cite{Catani:2000vq}.
The resummed and fixed-order 
procedures at small and large values of 
$q_T$ can then be matched at intermediate values of $q_T$,
to obtain QCD predictions for the entire range of transverse momenta.

In Refs.~\cite{Catani:2000vq,Bozzi:2005wk} we have proposed a method to 
perform transverse-momentum
resummation that introduces some novel features, and we briefly recall it below.

%%%%%%%%%%%

The resummation is performed at the level of the partonic cross section, which is decomposed as:
\begin{equation}
\label{resplusfin}
\f{d{\hat \sigma}_{F\,ab}}{dq_T^2}=
\f{d{\hat \sigma}_{F\,ab}^{(\rm res.)}}{dq_T^2}
+\f{d{\hat \sigma}_{F\,ab}^{(\rm fin.)}}{dq_T^2}\, .
\end{equation}
The first term on the right hand side,
$d{\hat \sigma}^{({\rm res.})}_{F\, ab}$,
contains all the logarithmically-enhanced contributions at small $q_T$,
and has to be evaluated by resumming them to all orders in $\as$.
The second term, $d{\hat \sigma}^{({\rm fin.})}_{F\, ab}$,
is free of such contributions, and
can thus be evaluated at fixed order in perturbation theory. 

The resummed component $d{\hat \sigma}^{({\rm res.})}_{F\, ab}$
can be expressed as
\begin{equation}
\label{resum}
\f{d{\hat \sigma}_{F \,ab}^{(\rm res.)}}{dq_T^2}(q_T,M,{\hat s};
\as(\mu_R^2),\mu_R^2,\mu_F^2) 
=\f{M^2}{\hat s} \;
\int_0^\infty db \; \f{b}{2} \;J_0(b q_T) 
\;{\cal W}_{ab}^{F}(b,M,{\hat s};\as(\mu_R^2),\mu_R^2,\mu_F^2) \;,
\end{equation}
where $J_0(x)$ is the $0$-order Bessel function, $\mu_R$ ($\mu_F$) is
the renormalization (factorization) scale and ${\hat s}$ is
the partonic centre-of-mass energy.
By taking the $N$-moments of ${\cal W}$ with respect to the variable $z=M^2/{\hat s}$ at fixed $M$
the resummation structure of ${\cal W}_{ab, \,N}^F$ can indeed be organized in exponential form.
\begin{align}
\label{wtilde}
{\cal W}_{N}^{F}(b,M;\as(\mu_R^2),\mu_R^2,\mu_F^2)
&={\cal H}_{N}^{F}\left(M, 
\as(\mu_R^2);M^2/\mu^2_R,M^2/\mu^2_F,M^2/Q^2
\right) \nonumber \\
&\times \exp\{{\cal G}_{N}(\as(\mu^2_R),L;M^2/\mu^2_R,M^2/Q^2
)\}
\;\;,
\end{align}
were we have defined the logarithmic expansion parameter $L$ as
\begin{equation}
\label{logpar}
L\equiv \ln \f{Q^2 b^2}{b_0^2}
\end{equation}
and the coefficient $b_0=2e^{-\gamma_E}$ ($\gamma_E=0.5772...$ is the Euler number) has a kinematical origin.
The scale $Q$ appearing in Eqs.~(\ref{wtilde},~\ref{logpar}),
named resummation scale in Ref.~\cite{Bozzi:2005wk},
parametrizes the
arbitrariness in the truncation of the resummation formula, 
and it has to be chosen of the order of the hard scale $M$.
Variations of $Q$ around $M$ can give an idea of the size of yet uncalculated
(or neglected) higher-order
logarithmic contributions.
The function ${\cal H}_N^{F}$ does not depend on the impact parameter $b$ and it includes all the perturbative
terms that behave as constants as $b\to\infty$. It can thus be expanded in powers of $\as=\as(\mu_R^2)$.
The exponent ${\cal G}_N$ includes the complete dependence on $b$ and, in particular, it contains all
the terms that order-by-order in $\as$ are logarithmically divergent as $b\to\infty$.

In the implementation of Eq.~(\ref{wtilde}) the resummation of the large logarithmic contributions affects
not only the small-$q_T$ region ($q_T\ll M$), but also the region of large $q_T$ ($q_T\sim M$).
This can be easily understood by observing that the logarithmic expansion parameter $L$ is divergent as $b\to 0$.
To reduce the impact of unjustified higher-order contributions in the large-$q_T$ region,
the logarithmic variable $L$ in Eq.~(\ref{logpar}) is replaced as
\begin{equation}
\label{ltilde}
L\to\tL~~~~~~~~~~~~\tL\equiv \ln \left(\f{Q^2 b^2}{b_0^2}+1\right)\, .
\end{equation}
The variables $L$ and $\tL$ are equivalent when $Qb\gg 1$, but they lead to a different behaviour
of the form factor at small values of $b$ (i.e. large values of $q_T$).
In fact, when $Qb\ll 1$, $\tL\to 0$ and $\exp\{{\cal G}_N\}\to 1$.
The replacement in Eq.~(\ref{ltilde}) has thus a twofold consequence: it reduces the impact of resummation at large values of $q_T$, and
it allows us to recover the corresponding fixed-order cross section upon integration over $q_T$.

Another important property of the formalism of Ref.~\cite{Bozzi:2005wk} is that
the process dependence (as well as the factorization scale and scheme dependence) is fully encoded
in the hard function ${\cal H}^{F}$.
In other words, the form factor $\exp\{{\cal G}_N\}$ is universal: it depends only on the channel
in which the process occurs at Born level ($q{\bar q}$ annihilation
in the case of vector-boson production, $gg$ fusion in the case of Higgs boson production). 
The explicit form of the universal
form factor is known up to next-to-leading logarithmic (NLL) 
\cite{Kodaira:1981nh,Catani:vd}
and next-to-next-to-leading
logarithmic (NNLL) \cite{Davies:1984hs,Davies:1984sp,deFlorian:2000pr} level.
The general form of the process dependent
hard function is known up to
the first relative order in $\as$ \cite{deFlorian:2000pr}.
The hard function has been computed up to the
second relative order in $\as$ in the cases of SM Higgs boson 
production \cite{Catani:2007vq} and DY lepton pair production \cite{Catani:2009sm}.
%=========================================
\begin{wrapfigure}{r}{0.5\columnwidth}
\centerline{\includegraphics[width=0.45\columnwidth]{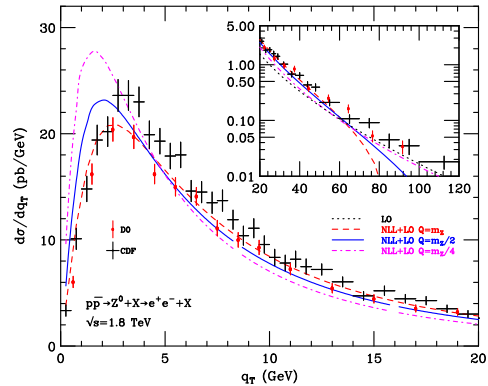}}
\caption{Z production at the Tevatron}\label{fig:ptZadep}
\end{wrapfigure}
%=========================================
The formalism that we have briefly recalled defines 
a systematic `order-by-order' (in extended sense)
expansion \cite{Bozzi:2005wk} of Eq.~(\ref{resplusfin}):
it can be used to obtain predictions that contain the full information of the
perturbative calculation up to a given fixed order plus resummation of
logarithmically-enhanced contributions from higher orders.
The various
orders of this expansion are denoted
as LL, NLL+LO, NNLL+NLO, etc., where the 
first label (LL, NLL, NNLL, $\dots$) refers to the logarithmic accuracy at 
small $q_T$ and the second label (LO, NLO, $\dots$) refers to the customary 
perturbative order
at large $q_T$. 
It is worthwhile noticing 
%Note 
that the NLL+LO (NNLL+NLO) result includes the {\em full} NLO (NNLO)
perturbative contribution in the small-$q_T$ region.
In particular,  the NLO (NNLO) result for total cross section  
is exactly recovered upon integration
over $q_T$ of the differential cross section $d \sigma/dq_T$ at NLL+LO 
(NNLL+NLO) accuracy.
%=========================================
\begin{wrapfigure}{r}{0.5\columnwidth}
\centerline{\includegraphics[width=0.45\columnwidth]{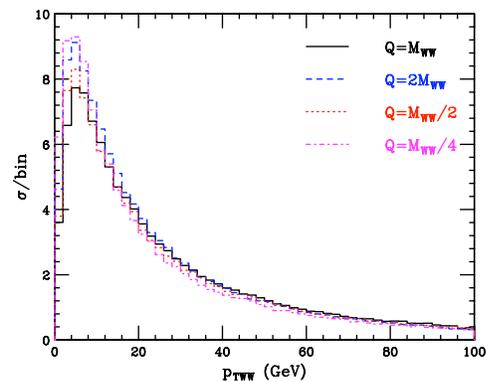}}
\caption{WW production at the LHC}\label{fig:ptWWadep}
\end{wrapfigure}
%=========================================

The method has so far been applied to SM Higgs boson 
production \cite{Bozzi:2003jy,Bozzi:2005wk,Bozzi:2007pn},
vector boson production \cite{Bozzi:2008bb},
$WW$ \cite{Grazzini:2005vw} and $ZZ$ \cite{Frederix:2008vb} pair production,
slepton and gaugino pair production \cite{Bozzi:2006fw,Debove:2009ia}.
In the following we review some of these results.
Other phenomenological studies of
$q_T$ resummation have been performed for vector boson production
\cite{Arnold:1990yk,Balazs:1995nz, Balazs:1997xd, Ellis:1997sc, Ellis:1997ii, Qiu:2000ga,Kulesza:2001jc, Kulesza:2002rh, Landry:2002ix, Berge:2005rv},
Higgs boson production \cite{Hinchliffe:1988ap,Kauffman:1991jt,Yuan:1991we,Balazs:2000wv,Balazs:2000sz,Berger:2002ut,Kulesza:2003wi,Kulesza:2003wn}, diphoton \cite{Balazs:1999yf} and $ZZ$ \cite{Balazs:1998bm} production.

We start from the case of vector boson production. In Ref.~\cite{Bozzi:2008bb}
we have presented NLL+LO results for $Z$ boson production at the Tevatron.
In Fig.~\ref{fig:ptZadep} we show the $q_T$ spectrum for different choices
of the resummation scale $Q$. The theoretical prediction appears to be in reasonably good agreement with the Tevatron data, but the uncertainty from missing higher-order logarithmic contributions is still large.
The extension of this calculation to NNLL+NLO is in progress and we expect
a considerable reduction of theoretical uncertainties.

%=========================================
\begin{wrapfigure}{r}{0.5\columnwidth}
\centerline{\includegraphics[width=0.45\columnwidth]{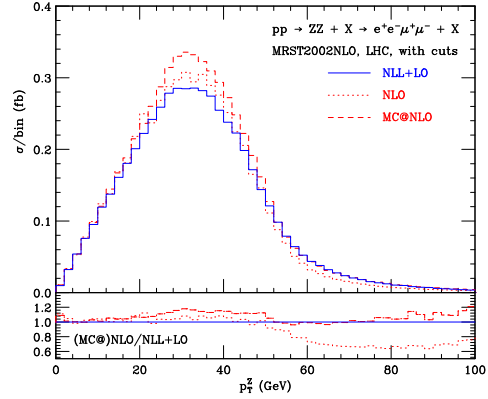}}
\caption{ZZ production at the LHC}\label{fig:ptZ_cuts}
\end{wrapfigure}
%=========================================
A similar picture emerges from the
work of Refs.~\cite{Grazzini:2005vw,Frederix:2008vb}.
Here the calculation includes the decay of the two vector bosons and full spin correlations.
In Fig.~\ref{fig:ptWWadep} \cite{Grazzini:2005vw}
we report the transverse momentum spectrum of the $WW$ pair at the LHC.
The uncertainties from missing higher order logarithmic contributions
are large and comparable to those of Fig.~\ref{fig:ptZadep}.
%=========================================
\begin{wrapfigure}{r}{0.5\columnwidth}
\centerline{\includegraphics[width=0.45\columnwidth]{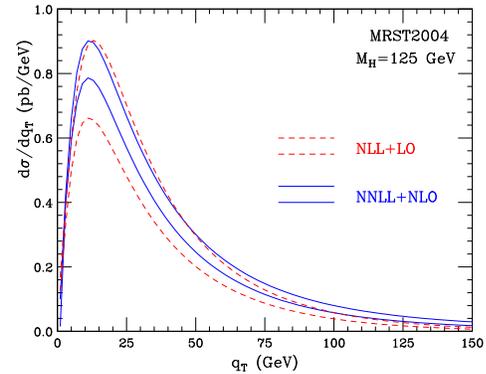}}
\caption{Higgs production at the LHC: NLL+LO and NNLL+NLO bands}\label{fig:bands}
\end{wrapfigure}
%=========================================
In Fig.~\ref{fig:ptZ_cuts} \cite{Frederix:2008vb} we show the impact of soft-gluon resummation on the spectrum of one of the two $Z$ bosons in the case of $ZZ$ production at the LHC. The result from MC@NLO \cite{MCatNLO} is also shown for comparison. The spectrum of the $Z$ boson is well behaved at NLO, but
the impact of the resummation is visible in its shape.
This is evident from the lower part of the plot, showing NLO and MC@NLO results normalized to NLL+LO.
The resummation effects, which are important for transverse-momentum spectra, are instead less relevant
for angular distributions.

In the case of the Standard Model Higgs boson, transverse-momentum resummation
has been performed at full NNLL+NLO accuracy. The calculation is implemented
in the program {\tt HqT} \cite{Bozzi:2005wk}.
In Fig.~\ref{fig:bands} the production of a Higgs boson of mass $m_H=125$ GeV at the LHC is considered,
and the NLL+LO and NNLL+NLO bands are compared.
The bands are obtained by varying $\mu_F$ and $\mu_R$ between $0.5 m_H$ and $2m_H$ with the constraint
$0.5\leq \mu_F/\mu_R\leq 2$.
The result shows that perturbative uncertainties are under control.
The uncertainty from missing higher-order logarithmic contributions,
estimated through resummation scale variations, are about $\pm 5\%$ at the peak.

I wish to thank my collaborators, Giuseppe Bozzi, Stefano Catani, Giancarlo Ferrera, Daniel de Florian and Rikkert Frederix, for a fruitful collaboration on the topics discussed in this contribution.

\begin{footnotesize}
% IF YOU DO NOT USE BIBTEX, USE THE FOLLOWING SAMPLE SCHEME FOR THE REFERENCES
% ----------------------------------------------------------------------------

\end{footnotesize}

\end{document}